\documentstyle[epsf]{article}

\newcommand{\ee}{\end{equation}}
\newcommand{\be}{\begin{equation}}
\def\adots{\cdot^{\displaystyle{~\cdot}^{\displaystyle{~\cdot}} }}
  
\begin{document}

\centerline {\bf\huge The Singularity of   }
\vskip 0.3 cm 
\centerline {\bf\huge Kontsevich's Solution for $QH^{*}(CP^2)$ }
\vskip 0.3 cm
\centerline{\bf   Davide Guzzetti}
\vskip 0.3 cm
\centerline{\bf  Research Institute for Mathematical Sciences (RIMS)}
\centerline{\bf  Kyoto University}
\centerline{\bf Kitashirakawa, Sakyo-ku, Kyoto 606-8502, Japan}
\vskip 0.2 cm 
\centerline{\bf Phone +81  75-753-7214, Fax +81 75-753-7266}
\centerline{\bf E-mail: guzzetti@kurims.kyoto-u.ac.jp}
\vskip 0.5 cm

\vskip 1 cm
\centerline{~~\bf Abstract.} 

 In this paper we study the nature of the singularity of the Kontsevich's 
 solution of the 
 WDVV equations of associativity.  We 
 prove that it corresponds to a 
singularity in the change of two coordinates systems of the Frobenius 
manifold given by 
the quantum cohomology of $CP^2$.
\vskip 0.3 cm

\section{Introduction}

 In this paper we study the nature of the singularity of the solution of the 
 WDVV equations of associativity for the quantum cohomology of the complex 
projective space $CP^2$. As we will explain in detail below, the {\it quantum cohomology} of a projective space $CP^d$ ($d$ integer) is a Frobenius Manifold 
which has a structure specified by a solution to a 
WDVV equation. In the case of $CP^2$ such a solution was
 found by Kontsevich  \cite{KM} in the form of a convergent series in the flat coordinates $(t^1,t^2,t^3)$ of the corresponding Frobenius manifold: 
\be 
  F(t):= {1\over 2} [(t^1)^2 t^3+t^1(t^2)^2)] + {1\over t^3} 
\sum_{k=1}^{\infty}  A_k \bigl[ (t^3)^3 \exp(t^2) \bigr]^k, 
~~~A_k \in {\bf R}   
 ~~~~~\label{baba}
\ee
 The series converges in a neighborhood of $(t^3)^3 \exp(t^2)=0$ with a 
certain radius of convergence extimated by Di Francesco and Itzykson  
\cite{DI}.  The coefficients $A_k$ are real and are the Gromov-Witten invariants of genus zero. We will explain this point later. 
As for the Gromov-Witten invariants of genus one of $CP^2$, 
we refer to \cite{DZ}, where  
 B. Dubrovin and Y. Zhang  proved that their 
$G$-function has the same radius of convergence of (\ref{baba}). 

As we will explain below, 
the nature of the boundary points of the ball of convergence of (\ref{baba}) 
 is important to study of the global structure of the  manifold.

 In the following, we first state rigorously the problem of 
 the global structure of a Frobenius manifold, then we introduce 
 the quantum cohomology of $CP^d$ as a Frobenius manifold and we explain 
its importance in enumerative geometry. 
 Finally, we study the boundary points 
of the ball of convergence of Kontsevich's solution. 
We prove that they  correspond to a 
singularity in the change of two coordinates systems. 
 
 Our paper is part of a project to 
study of the global structure of  Frobenius manifolds 
 that we started in \cite{guz1}. 

\vskip 0.3 cm

\subsection{Frobenius manifolds and their global structure} 
 The subject of this subsection can be found in \cite{Dub1}, \cite{Dub2} or, 
in a more synthetic way, in \cite{guz1}. 
  
The WDVV equations of associativity were introduced by Witten \cite{Witten},
Dijkgraaf, Verlinde E., Verlinde H. \cite{DVV}.  
They are  
differential equations satisfied by the {\it primary free energy} $F(t)$ in 
 two-dimensional topological field theory. $F(t)$  is a function 
of the coupling constants $t:=(t^1,t^2,...,t^n)$ $t^i\in {\bf C}$. Let
$\partial_{\alpha}:={\partial \over \partial t_{\alpha}}$.   Given a 
non-degenerate  symmetric matrix $\eta^{\alpha \beta}$, $\alpha,\beta=1,...,n$,
 and numbers  $q_1,q_2,...,q_n$, $r_1,r_2,...,r_n$,  $d$, ($r_{\alpha}=0$ if
$q_{\alpha}\neq 1$, $\alpha=1,...,n$), 
  the WDVV equations are
\be 
\partial_{\alpha}\partial_{\beta} \partial_{\lambda} F~\eta^{\lambda \mu} 
~\partial_{\mu}\partial_{\gamma}\partial_{\delta} F ~=~ \hbox{ the same with 
$\alpha$, $\delta$ exchanged},
\label{WDVV1}
\ee
\be
\partial_1 \partial_{\alpha}\partial_{\beta} F = \eta_{\alpha\beta},
\label{WDVV2}
\ee
\be
  E (F)= (3-d) F + \hbox{ (at most) quadratic terms}
\label{WDVV3b},
\ee
where the matrix $
 (\eta_{\alpha \beta})$ is the inverse of the matrix $(\eta^{\alpha \beta})$ 
and the differential operator $E$ is $
E:= \sum_{\alpha=1}^n E^{\alpha} \partial_{\alpha}$, $E^{\alpha}:=
       (1-q_{\alpha})t^{\alpha} +r_{\alpha}$, $\alpha=1,...,n$,  
and will be called {\it Euler vector field}.

  Frobenius structures first appeared in the works of K. Saito 
\cite{Saito} \cite{SYS} with the name of {\it flat scructures}.  The complete 
theory of Frobenius manifolds was then developed by 
 B. Dubrovin  as a geometrical setting for topological field 
theory and WDVV equations in \cite{Dub4}. Such a theory 
 has  links to many branches of
mathematics  like singularity theory and 
 reflection groups \cite{Saito} \cite{SYS}  \cite{Dub6}  \cite{Dub1},
algebraic and 
enumerative geometry \cite{KM} \cite{Manin}, isomonodromic deformations
theory, boundary value problems 
 and Painlev\'e equations \cite{Dub2}.  

If we define $ c_{\alpha \beta \gamma}(t):= \partial_{\alpha}\partial_{\beta} 
\partial_{\gamma} F(t)$, $c_{\alpha\beta}^{\gamma}(t) := \eta^{\gamma \mu} 
c_{\alpha \beta \mu}(t)$   (sum over repeated indices is always omitted in the 
paper), 
and we consider a vector space $A$=span($e_1,...,e_n$), then we obtain a 
family of commutative algebras $A_t$  with the  
multiplication 
$ 
  e_{\alpha} \cdot e_{\beta} := c_{\alpha \beta}^{\gamma}(t) e_{\gamma}. 
$ 
Equation (\ref{WDVV1}) is equivalent to associativity and (\ref{WDVV2}) 
implies that $e_1$ is the unity.

\vskip 0.2 cm 
\noindent
{\bf Definition:} 
 A {\it Frobenius manifold} 
is a smooth/analytic manifold $M$ 
over ${\bf C}$ whose tangent space $T_t M$ at any $t \in M$ 
is an {\it associative, commutative algebra} with {\it unity} $e$. Moreover,
there  
exists a non-degenerate bilinear form $<~~~, ~~~>$ 
 defining a {\it flat metric} (flat means that the curvature associated to the Levi-Civita connection is zero).

\vskip 0.2 cm 

 We denote by $\cdot$ the product and by $\nabla$  the covariant derivative 
of $<\cdot,\cdot>$. We require that the tensors  
$c(u,v,w):=<u\cdot v , w>$, and $\nabla_y c(u,v,w)$,  $u,v,w,y\in T_t M$,  be  
symmetric. 
Let 
$t^1,..,t^n$ be (local)  flat coordinates for $t\in M$. Let  
$e_{\alpha}:=\partial_{\alpha}$ be the canonical basis 
in $T_t M$,  $\eta_{\alpha \beta}:= <\partial_{\alpha},
\partial_{\beta}>$,
 $ c_{\alpha \beta \gamma}(t):=<\partial_{\alpha} \cdot \partial_{\beta},
\partial_{\gamma}>$. The symmetry of $c$ corresponds to 
the complete symmetry of $\partial_{\delta}
c_{\alpha \beta \gamma}(t)$ in the indices. This implies the existence 
of a function $F(t)$ such that $\partial_{\alpha}\partial_{\beta}
\partial_{\gamma} F(t) 
= c_{\alpha \beta \gamma}(t)$ satisfying the WDVV (\ref{WDVV1}). The 
equation  
(\ref{WDVV2}) follows from  the axiom $\nabla e=0$ which yields
 $e=\partial_1$. . Some more axioms are needed to formulate the 
quasi-homogeneity condition (\ref{WDVV3b}) and we 
refer the reader to \cite{Dub1} \cite{Dub2} \cite{Dub3}.   
  In this way the WDVV equations are reformulated in a geometrical terms. 

\vskip 0.2 cm 

We first 
consider the problem of the local structure of Frobenius manifolds.   
  A Frobenius manifold is characterized by a family of 
 {\it flat} connections $\tilde{\nabla}(z)$  
 parameterized by a complex number $z$, such that for $z=0$
 the connection is associated to $<~~,~~>$. For this reason  $\tilde{\nabla}(z)$ are  called {\it deformed connections}.  Let $u,v\in T_tM$, ${d\over dz} \in 
T_z {\bf C}$;  the 
family  is defined on $M\times {\bf C}$ as follows:
$$ 
    \tilde{\nabla}_u v := \nabla_u v + z u \cdot v,$$ 
$$
   \tilde{\nabla}_{d\over dz}v:= {\partial \over \partial z} v +E\cdot v - 
{1\over z} \hat{\mu}v,
$$ 
$$ 
  \tilde{\nabla}_{d\over dz}{d\over dz} =0,~~~~~~~
 \tilde{\nabla}_u {d\over dz} =0
$$
where 
  $E$ is the Euler vector field and 
$$
\hat{\mu}:= I-{d \over 2} -
\nabla E
$$ 
is an operator acting on $v$. In flat coordinates $t=(t^1,...,t^n)$, 
 $\hat{\mu}$ becomes: 
$$
\hat{\mu} = \hbox{diag}(\mu_1,...,\mu_n),~~~~\mu_{\alpha}=q_{\alpha} -{d 
\over 2},$$ provided that $\nabla E$ is diagonalizable. This will be assumed in the paper.  
A flat coordinate $\tilde{t}(t,z)$ is a solution of 
 $\tilde{\nabla} d\tilde{t}
=0$, which is a linear system
\be
    \partial_{\alpha} \xi = z C_{\alpha}(t) \xi,
\label{systemt}
\ee
\be
     \partial_z \xi = \left[{\cal U}(t) +{\hat{\mu} \over z} \right] \xi,
\label{systemz}
\ee
where 
$ 
   \xi$ is a column vector of components $\xi^{\alpha}=\eta^{\alpha\mu}
{\partial \tilde{t} \over\partial t^{\mu}}$, $\alpha=1,...,n$  
and $ C_{\alpha}(t):=
\bigl(
                                  c_{\alpha \gamma}^{\beta}(t)\bigr)$, $
{\cal U}:= \bigl( E^{\mu} c_{\mu \gamma}^{\beta}(t)  \bigr)$ are $n\times n$ 
matrices.

\vskip 0.2 cm 
The quantum cohomology of projective spaces, 
to be introduce below, 
belongs to the class of  {\it semi-simple} Frobenius manifolds, 
namely analytic Frobenius manifolds such that the 
matrix ${\cal U}$ can be diagonalized with {\it distinct eigenvalues} on an 
open dense subset ${\cal M}$
 of $M$. 
Then, there exists an invertible matrix $\phi_0=\phi_0(t)$ such that 
$
   \phi_0 {\cal U} \phi_0^{-1} = \hbox{diag} (u_1,...,u_n)=:U$, $u_i\neq u_j 
$ for  $i\neq j$ 
 on ${\cal M}$. The systems (\ref{systemt}) and (\ref{systemz})  become:
\be
{\partial y \over \partial u_i}=\left[ zE_i+V_i\right]~y
\label{systemt1}
\ee
\be
{\partial y \over \partial z} = \left[U +{V\over z}\right] y,
\label{systemz1}
\ee
where the row-vector $y$ is $y:=\phi_0 ~\xi$,  $E_i$ is a 
diagonal matrix such that
$(E_i)_{ii}=1$ and $(E_i)_{jk}=0$ otherwise, and 
  $$
V_i:= {\partial
\phi_0\over \partial u_i}~\phi_0^{-1}
~~~~~V:=\phi_0 ~\hat{\mu}~ \phi_0^{-1}, 
$$

As it is proved in \cite{Dub1} \cite{Dub2}, 
 $u_1$,...,$u_n$ are local coordinates on ${\cal M}$. The two bases ${\partial \over \partial t^{\nu}}$, $\nu=1,...,n$ and ${\partial \over \partial u_i}$, $i=1,...,n$ are related by $\phi_0$ according to the linear combination ${\partial \over \partial t^{\nu}}= \sum_{i=1}^n {(\phi_0)_{i\nu}\over (\phi_0)_{i1}} ~{\partial \over \partial u_i}$. 
 Locally we obtain a  change of coordinates, $t^{\alpha} = 
t^{\alpha}(u)$, then $\phi_0=\phi_0(u)$, 
$V=V(u)$. The local Frobenius structure of ${\cal M}$ 
is given  by  
parametric formulae: 
\be
t^{\alpha}=t^{\alpha}(u),~~~~~F=F(u)
\label{parametricintroduzione}
\ee
where $t^{\alpha}(u)$, $F(u)$ are certain meromorphic functions  of
$(u_1,...,u_n)$, $u_i\neq u_j$, which  can be obtained from 
$\phi_0(u)$ and $V(u)$.  Their explicit construction
was the object of \cite{guz1}. We stress here that the condition $u_i\neq u_j$ is crucial. We will further
 comment on this when we face the problem of the global structure.

\vskip 0.2 cm 

  The dependence of the system  on $u$ is {\it
  isomonodromic}. 
This means that the monodromy data of the system (\ref{systemz1}), to
  be  introduced below, do not change for a small
deformation of $u$.   Therefore, the coefficients of the system 
 in every
local chart of ${\cal M}$   are 
 naturally labeled by the monodromy data. To calculate the functions
  (\ref{parametricintroduzione})  in every local chart one has to reconstruct
  the  system  ({\ref{systemz1})  
from its {\it monodromy data}. This is  the 
{\it inverse problem}.

\vskip 0.3 cm 

  We briefly explain what are the monodromy data of the system (\ref{systemz1}) and why they do not depend on $u$ (locally). For details the reader is referred to \cite{Dub2}. At $z=0$ the system  (\ref{systemz1}) has a fundamental matrix solution (i.e. an invertible $n\times n$ matrix solution) of the form 
 \be 
  Y_0(z,u)= \left[\sum_{p=0}^{\infty}\phi_p(u)~z^p \right] 
z^{\hat{\mu}} z^{R},
\label{Y0}
\ee
  where $ {R}_{\alpha \beta}=0$  if $ \mu_{\alpha}-\mu_{\beta} 
\neq k>0$, $k\in {\bf N}$. 
At $z=\infty$ there is a formal $n\times n$ matrix 
 solution  of (\ref{systemz1}) given by 
$
   {Y}_F = \left[ I + {F_1(u) \over z}+{F_2(u) \over z^2}+
... \right]~e^{z~U}$ 
where $F_j(u)$'s are $n \times n$ matrices. It is a well known   result that 
there exist  fundamental matrix  solutions with asymptotic expansion  
${Y}_F$ as 
  $z \to \infty$ \cite{BJL1}. 
 Let $l$ be a 
  generic oriented line  passing through the origin.  Let  $l_{+}$ be the
 positive half-line and  $l_{-}$ the negative one. 
 Let  $\Pi_L$ and $\Pi_R$ be two sectors in the complex plane 
to the
 left and to the right of $l$ respectively. 
 There exist unique fundamental matrix solutions  $Y_L$ and $Y_R$ having the 
asymptotic expansion $Y_F$ for $x\to \infty$ 
in $\Pi_L$ and $\Pi_R$ 
respectively \cite{BJL1}.  They are related by an invertible 
 connection matrix $S$,
called {\it 
Stokes matrix}, such that $
    {Y}_L(z)={Y}_R(z)S
$  
for $z\in l_{+}$. As it is proved in \cite{Dub2} we also 
have $Y_L(z)=Y_R(z) S^T$ 
on $l_{-}$. 
  
Finally, 
there exists a $n \times n$ invertible {\it  connection matrix} 
 $C$ such that $Y_0=Y_R C$ on $\Pi_R$. 

\vskip 0.2 cm 
\noindent 
  {\bf Definition:} 
The matrices $R$, $C$,  $\hat{\mu}$ and the Stokes matrix $S$ of the system
(\ref{systemz1}) are the  {\it monodromy data} of the Frobenius
manifold 
 in a neighborhood of the point $u=(u_1,...,u_n)$. It is also necessary 
to specify which is the first eigenvalue of $\hat{\mu}$, because the 
dimension of the manifold is $d=-2\mu_1$  
(a more precise definition of monodromy data is in \cite{Dub2}).

\vskip 0.2 cm 
The definition makes sense because the data 
 do not change if $u$ undergoes a small
deformation. This problem is discussed in \cite{Dub2}. We also refer the reader to \cite{JMU} for a general discussion of isomonodromic deformations. Here we just observe that since a fundamental matrix solution $Y(z,u)$ 
 of (\ref{systemz1}) also satisfies (\ref{systemt1}), then the monodromy data 
  can not depend on $u$ (locally). In fact, 
${\partial Y \over \partial u_i} Y^{-1}= z E_i +V_i$ is single-valued 
in $z$.

\vskip 0.3 cm

The inverse problem can be formulated as a 
 {\it boundary value problem} (b.v.p.).  
 Let's fix $u=u^{(0)}=(u_1^{(0)},...,u_n^{(0)})$ such that $u_i^{(0)}\neq
u_j^{(0)} $ for $i\neq j$.  
Suppose we give $\mu$, $\mu_1$,  $R$, 
an admissible line $l$, $S$ and $C$.   
Some more technical conditions must be added, but we refer to \cite{Dub2}. 
Let $D$ be a disk specified by $|z|<\rho$ for some small $\rho$. Let $P_L$ and
$P_R$ be the intersection of the complement  of the disk with $\Pi_L$ and
$\Pi_R$ respectively. We denote by $\partial D_R$ and $\partial D_L$  the lines
 on the boundary of 
$D$ on the side of $P_R$ and $P_L$ respectively; we denote by $\tilde{l}_{+}$
and $\tilde{l}_{-}$
 the portion of $l_{+}$ and $l_{-}$  
on the common boundary of $P_R$ and $P_L$.
 Let's consider the following discontinuous b.v.p.: we want to 
  construct a piecewise holomorphic $n\times n$ matrix function 
$$
   \Phi(z)=\left\{\matrix{\Phi_R(z),~~~z\in P_R\cr
                           \Phi_L(z),~~~z\in P_L\cr
\Phi_0(z),~~~z\in D\cr
}\right.,
$$
continuous on the boundary of $P_R$, $P_L$, $D$ respectively, such that 
$$
   \Phi_L(\zeta)=\Phi_R(\zeta)~e^{\zeta U}S e^{-\zeta U},~~~~\zeta\in
   \tilde{l}_{+}
$$
$$
  \Phi_L(\zeta)=\Phi_R(\zeta)~e^{\zeta U}S^T e^{-\zeta U},~~~~\zeta\in
   \tilde{l}_{-}
$$
$$
\Phi_0(\zeta)=\Phi_R(\zeta)~e^{\zeta U} C\zeta^{-R}\zeta^{-\hat{\mu}}
,~~~~\zeta\in \partial D_R
$$
$$
\Phi_0(\zeta)=\Phi_L(\zeta)~e^{\zeta U}S^{-1} C\zeta^{-R}\zeta^{-\hat{\mu}}
,~~~~\zeta\in \partial D_L
$$
$$
   \Phi_{L/R}(z)\to I \hbox{ if  $z\to \infty$ in $P_{L/R}$ }. 
$$
 The reader may observe that 
 $ \tilde{Y}_{L/R}(z):=\Phi_{L/R}(z) e^{zU},
$  
$
    \tilde{Y}^{(0)}(z):=\Phi_0(z,u)z^{\hat{\mu}} z^{R}$
 have precisely the 
monodromy
 properties of the
 solutions of (\ref{systemz1}). 

\vskip 0.2 cm
\noindent
{\bf Theorem} \cite{Miwa}\cite{Malgrange}\cite{Dub2}: {\it If the above boundary value problem has solution for a given
$u^{(0)}= (u_1^{(0)},...,u_n^{(0)})$ such that $u_i^{(0)}\neq
u_j^{(0)} $ for $i\neq j$, then: 

i)  it is unique. 

ii) The solution exists and it is analytic 
for $u$ in a neighborhood of $u^{(0)}$. 

iii) The solution has analytic continuation as a meromorphic function 
on the universal covering of ${\bf C}^n\backslash\{diagonals\}$, 
where ``diagonals'' stands for the union of all the 
sets $\{ u \in {\bf C}^n ~|~ u_i=u_j, ~~i\neq j\}$. 
}

\vskip 0.2 cm
A  solution  $\tilde{Y}_{L/R}$, $\tilde{Y}^{(0)}$ of the b.v.p.
 solves the system (\ref{systemt1}), (\ref{systemz1}).  
This means that we can locally reconstruct 
  $V(u)$, $\phi_0(u)$  and (\ref{parametricintroduzione}) 
from the local solution of the b.v.p.. It follows that every local chart 
of the atlas covering the manifold is labeled by monodromy data. 
Moreover, $V(u)$, $\phi_0(u)$  and (\ref{parametricintroduzione}) 
can be continued analytically as 
meromorphic functions on the universal covering of ${\bf C}^n \backslash 
\hbox{diagonals}$.

Let  ${\cal S}_n$ be the symmetric group of $n$ 
elements.  Local 
coordinates $(u_1,...,u_n)$ are defined up to permutation. Thus, the 
analytic continuation of the local structure of  ${\cal M}$ 
is described by the 
{\it braid group} ${\cal B}_n$, namely the 
fundamental group of 
$({\bf C}^n \backslash 
\hbox{diagonals})/ {\cal S}_n $. There exists an action of the 
braid group itself on the monodromy data,  
 corresponding to the change of 
coordinate chart. 
 The group is generated by $n-1$ elements $\beta_1$,...,$\beta_{n-1}$  
such that 
$\beta_i$   is  represented  as a  
 deformation 
consisting  of  a permutation of $u_i$, $u_{i+1}$ moving 
counter-clockwise (clockwise or counter-clockwise is a matter of
convention).

 If $u_1$, ..., $u_n$ are  in
lexicographical order w.r.t. $l$, so that $S$ is upper
triangular,  the braid $\beta_i$ acts on $S$ as follows \cite{Dub2}: 
$$ 
   S \mapsto S^{\beta_i}= A_{i}(S)~S~A_{i}(S)
$$
where 
$$ 
   \left(A_{i}(S) \right)_{kk}=
   1~~~~~~k=1,...,n~~~n\neq~i,~i+1
$$
$$
 \left(A_{i}(S) \right)_{i+1,i+1}=-s_{i,i+1}
$$
$$
 \left(A_{i}(S) \right)_{i,i+1}=
 \left(A_{i}(S) \right)_{i+1,i}=1
$$
and all the other entries are zero. For a generic braid $\beta$ the action $S 
\to S^{\beta}$ is decomposed into a sequence of elementary transformations as above.  
 In this way, we are able to describe the analytic continuation of the local structure  in terms of monodromy data.

\vskip 0.2 cm
 Not all the braids are actually to be considered.  Suppose we 
do the following gauge  $y\mapsto J y$, $J=$diag$(\pm1,...,\pm1)$,
on the system (\ref{systemz1}).  Therefore ${J}U{J}^{-1}\equiv U$
but $S$ is transformed to  ${J} S {J}^{-1}$, where some  entries
change sign. The formulae  
which define a local chart of the manifold in
terms of monodromy data, which are given in \cite{Dub2}, \cite{guz1},   
 are not affected by this transformation.  The analytic continuation of the
local structure on the universal covering  of $({\bf
C}^n\backslash\hbox{diagonals})/{\cal S}_n$ 
  is therefore described by the elements of  the quotient group 
\be
   {\cal B}_n/ \{\beta\in {\cal B}_n ~|~S^{\beta}=JSJ\}
\label{quozienteinutile}
\ee
From these considerations it is proved in  \cite{Dub2} that:

\vskip 0.2 cm
\noindent 
{\bf Theorem} \cite{Dub2}: 
{\it Given monodromy data ($\mu_1$, $\hat{\mu}$, $R$, $S$, $C$), the
local Frobenius structure
obtained from the solution of the b.v.p. extends to an open 
dense  subset of the covering of  $({\bf
C}^n\backslash\hbox{diagonals})/{\cal S}_n$ w.r.t. the covering
transformations  (\ref{quozienteinutile}).}
\vskip 0.2 cm
{\it Let's  start from a Frobenius manifold $M$ of dimension $d$. 
Let ${\cal M}$ be the open
sub-manifold where  ${\cal U}(t)$ has distinct eigenvalues. If
we  compute its monodromy
data   ($\mu_1=-{d\over 2}$, $\hat{\mu}$, $R$, $S$, $C$) at a point $u^{(0)}\in {\cal M}$ 
and we construct the Frobenius structure from the
analytic continuation of the corresponding b.v.p. on the covering of  $({\bf
C}^n\backslash\hbox{diagonals})/{\cal S}_n$ w.r.t. the quotient
(\ref{quozienteinutile}), then there is
an equivalence of Frobenius structures between this last manifold and ${\cal
M}$. 
}

\vskip 0.3 cm 
 To understanding the {\it global structure} of a 
Frobenius manifold  we have to
study (\ref{parametricintroduzione}) 
when two  or more distinct coordinates $u_i$, $u_j$, etc,  merge. $\phi_0(u)$, 
$V(u)$ and 
 (\ref{parametricintroduzione}) are multi-valued meromorphic functions of
$u=(u_1,...,u_n)$ and the branching occurs when $u$ goes around a 
 loop around
the set of diagonals $\bigcup_{ij}\{ u \in {\bf C}^n ~|~ u_i=u_j, ~~i\neq
j\}$.    $\phi_0(u)$, $V(u)$ and 
 (\ref{parametricintroduzione}) have singular behavior if $u_i\to u_j$
$(i\neq j$). We call such
behavior {\it critical behavior}.

The Kontsevich's solution introduced at the beginning has a radius of convergence 
 which might be due to the fact that some coordinates $u_i$, $u_j$ merge at the boundary of the ball of convergence. 
We will prove that this is not the case. 
Rather, there is  a singularity in the change of 
coordinates $u \mapsto t$. 


\subsection{ Intersection Form  of a Frobenius Manifold}

 The deformed flat connection was introduced as a natural structure on a
 Frobenius manifold and allows to transform the problem of solving the WDVV
 equations to a problem of isomonodromic deformations. There is a further
 natural structure on a Frobenius manifold which makes it possible to do the
 same. It is the intersection form. We need it as a tool to calculate the canonical coordinates later. 

\vskip 0.2 cm
There is a natural isomorphism $\varphi:T_tM \to T_t^{*}M$ induced by
$<.,.>$. Namely,  
let $v\in T_tM$ and define $\varphi(v):=<v,.>$. This allow us to define the
product in $T_t^{*}M$ as follows: for $v,w\in T_tM$ we define $\varphi(v)\cdot
\varphi(w):=<v\cdot w,.>$. In flat coordinates $t^1,...,t^n$ the product is 
$$
    dt^{\alpha}\cdot dt^{\beta}=c^{\alpha\beta}_{\gamma}(t)~dt^{\gamma},~~~~
c^{\alpha\beta}_{\gamma}(t)=\eta^{\beta\delta}c_{\delta\gamma}^{\alpha}(t),
$$
(sums over repeated indices are omitted).

\vskip 0.2 cm
\noindent
{\bf Definition:} The {\it intersection form} at $t\in M$ is a bilinear form on
$T_t^{*}M$ defined by
$$
   (\omega_1,\omega_2):=(\omega_1\cdot\omega_2) (E(t))
$$
where $E(t)$ is the Euler vector field. In coordinates 
$$
g^{\alpha\beta}(t):=
(dt^{\alpha},dt^{\beta})= E^{\gamma}(t) c_{\gamma}^{\alpha \beta}.
$$

 In the semi-simple case, let $u_1,...,u_n$ be local canonical coordinates,
equal to the distinct 
eigenvalues of ${\cal U}(t)$. From the definitions we have  
$$
du_i\cdot du_j= {1\over \eta_{ii}}\delta_{ij} du_i,~~~
   g^{ij}(u)= (du_i,du_j)= {u_i\over \eta_{ii}} \delta_{ij},
   ~~~~~\eta_{ii}=(\phi_0)_{i1}^2 
$$ 
Then $g^{ij}-\lambda\eta^{ij}= {u_i-\lambda\over \eta_{ii}} \delta_{ij}$ and 
$$ 
  \det((g^{ij}-\lambda\eta^{ij}))= {1\over
  \det((\eta_{ij}))}~(u_1-\lambda)(u_2-\lambda)...(u_n-\lambda).
$$ 
Namely, the roots $\lambda$ of the above polynomial are the canonical
coordinates. 

In order to compute  $g^{\alpha\beta}$, in the paper we are going to use the 
following formula.  We
differentiate twice the 
expression 
$$
     E^{\gamma}\partial_{\gamma}F= (2-d)F +{1\over 2} A_{\alpha\beta}t^{\alpha}
     t^{\beta}+ B_{\alpha} t^{\alpha} +C
$$ 
which is the  quasi-homogeneity of $F$ up to quadratic terms. By recalling that 
$E^{\gamma}= (1-q_{\gamma})t^{\gamma}+r_{\gamma}$ and that
$\partial_{\alpha}\partial_{\beta} \partial_{\gamma} F=c_{\alpha\beta\gamma}$
we obtain 
\be 
   g^{\alpha\beta}(t)=
   (1+d-q_{\alpha}-q_{\beta})\partial^{\alpha}\partial^{\beta} F(t)
   +A^{\alpha\beta} 
\label{dall'MSRI}
\ee
where $\partial^{\alpha}=\eta^{\alpha \beta}\partial_{\beta}$, $A^{\alpha
\beta} = \eta^{\alpha \gamma}\eta^{\beta\delta} 
A_{\gamma\delta} $.



\section{ Quantum Cohomology of Projective spaces}\label{cap2}

 In this section we introduce the Frobenius manifold called quantum
 cohomology of the projective space $CP^d$ and we describe its connections to
 enumerative geometry. 

It is possible to introduce a
 structure of Frobenius algebra on the cohomology $H^{*}(X,{\bf C})$ of a
 closed oriented manifold $X$ of dimension $d$  such that 
$$ 
    H^{i}(X,{\bf C})=0 ~~\hbox{ for $i$ odd}.
$$
Then
$$ 
    H^{*}(X,{\bf C})=\otimes_{i=0}^d ~H^{2i}(X,{\bf C}).
$$
For brevity we omit ${\bf C}$ in $H$. 
$H^{*}(X)$ can be realized by classes of closed differential forms. The
unit element is a 0-form $e_1\in  H^{0}(X)$. Let us  denote by 
$\omega_{\alpha}$ a form in 
$H^{2q_{\alpha}}(X)$, where $q_1=0$, $q_2=1$, ...,
$q_{d+1}=d$. The product of two forms $\omega_{\alpha}$, $\omega_{\beta}$  
 is defined by the wedge product  $ \omega_{\alpha} \wedge\omega_{\beta}\in
H^{2(q_{\alpha}+q_{\beta})}(X)$ and  
the
 bilinear form is 
$$ 
  <\omega_{\alpha},\omega_{\beta}>:=\int_X ~ \omega_{\alpha}
  \wedge\omega_{\beta} \neq 0 ~\Longleftrightarrow ~q_{\alpha}+q_{\beta}=d
$$
It is not degenerate by Poincar\'e duality  and    
$q_{\alpha}+q_{d-\alpha+1}=d$.

\vskip 0.3 cm

Let $X=CP^d$. 
Let $e_1=1\in H^0(CP^d)$, $ e_2\in H^2(CP^d)$,  ...,
 $e_{d+1}\in H^{2d}(CP^d)$ be a basis in $ H^{*}(CP^d)$. For a suitable
normalization we have 
$$ 
   (\eta_{\alpha\beta}):=(<e_{\alpha},e_{\beta}>)= 
\pmatrix{             &   &         & & 1 \cr 
                      &   &         &1& \cr 
                      &   & \adots  & & \cr
                      & 1 &         & & \cr
                    1 &   &         & &  \cr
                   } 
$$
The multiplication is $$e_{\alpha}\wedge e_{\beta}= e_{\alpha+\beta-1}.$$ We
observe that it can also be written as 
$$ 
  e_{\alpha}\wedge e_{\beta}=c_{\alpha\beta}^{\gamma} e_{\gamma} ,~~\hbox
  { sums on $\gamma$} 
$$
where 
    $$\eta_{\alpha\delta}c_{\beta\gamma}^{\delta}:= {\partial^3 F_0(t) \over 
   \partial t^{\alpha}  \partial t^{\beta}  \partial t^{\gamma}}
$$
$$
      F_0(t):= {1\over 2} (t^1)^2 t^n +{1\over 2} t^1 \sum_{\alpha=2}^{n-1}
  t^{\alpha} t^{n-\alpha+1}
$$
$F_0$ is the trivial solution of WDVV equations. We can construct a trivial
Frobenius manifold whose points are $t:=\sum_{\alpha=1}^{d+1} t^{\alpha}
e_{\alpha}$. It has tangent space $H^{*}(CP^d)$ at any $t$. By {\it quantum
cohomology} of $CP^d$ (denoted by $QH^{*}(CP^d)$)  
we mean a Frobenius manifold whose structure is
specified by 
$$
    F(t)=F_0(t)+\hbox{ analytic perturbation}
$$
 This manifold has therefore tangent spaces $T_tQH^{*}(CP^d)= H^{*}(CP^d)$,
 with the same $<.,.>$ as above, 
 but the multiplication is a deformation, depending on $t$, 
of the wedge product (this is the
 origin of the adjective ``quantum'').

\section{ The case of $CP^2$}
 
 To start with, we  restrict to  $CP^2$. In this case 
$$ 
  F_0(t)={1\over 2}\left[ (t^1)^2 t^3+t^1(t^2)^2\right]
$$ 
which generates the product for the basis $e_1=1\in H^0$, $e_2\in H^2$,
$e_3\in H^4$. The deformation was introduced by Kontsevich \cite{KM}.

\subsection{Kontsevich's solution}

 The WDVV equations for $n=3$ variables have  solutions  
 $$F(t_1, t_2,t_3)= F_0(t_1,t_2,t_3)+f(t_2,t_3).$$
 $f(t_2,t_3)$ satisfies a differential equation obtained 
by substituting $F(t)$ into the WDVV equations. Namely: 
\be
   f_{222}f_{233}+f_{333}=(f_{223})^2 
\label{WDVV1special}
\ee
with the notation $f_{ijk}:={\partial^3 f \over \partial t_i \partial t_j 
\partial t_k}$. As for notations, the variables $t_j$ are flat coordinates in  
the Frobenius manifold associate to $F$. They should be written with upper 
indices, but we use the lower for convenience of notation. 

Let $N_k$ be the number of rational curves $CP^1 \to CP^2$ of degree $k$ 
through $3k-1$ generic points. Kontsevich \cite{KM} constructed the solution
\be
  f(t_2,t_3) = {1\over t_3} \varphi(\tau),~~~~
\varphi(\tau)=\sum_{k=1}^{\infty} A_k \tau^k,~~~
~~\tau= t_3^3~e^{t_2}
\label{Konts}
\ee
 where
 $$
 A_k= {N_k\over (3k-1)!}
$$
 The $A_k$ (or $N_k$) are called Gromov-Witten invariants of genus zero. 
We note that this solution has precisely the form of the general solution of
 the WDVV eqs. for $n=3$, $d=2$ and $r_2=3$ \cite{Dub1}. 
  If we put $\tau=e^X$ and we define 
$$
\Phi(X):=\varphi(e^X)=\sum_{k=1}^{\infty} ~A_k ~e^{kX},
$$
we  rewrite (\ref{WDVV1special}) as follows: 
\be
    -6 \Phi+33\Phi^{\prime} -54 \Phi^{\prime \prime} -  
(\Phi^{\prime \prime})^2
      + \Phi^{\prime \prime \prime} \left(
             27+2 \Phi^{\prime} -3 \Phi^{\prime \prime}
             \right)  =0
\label{WDVVK}
\ee
 The prime stands for the derivative w.r.t $X$. 
If we fix $A_1$, the above (\ref{WDVVK}) determines the $A_k$ uniquely. Since 
$N_1=1$, we fix 
$$A_1={1\over 2}.
$$
Then  (\ref{WDVVK})  yields  the recurrence relation
\be
  A_k= \sum_{i=1}^{k-1} \left[ {A_i A_{k-i} ~i(k-i)\bigl(
                                (3i-2)(3k-3i-2)(k+2)+8k-8
\bigr)\over 6(3k-1)(3k-2)(3k-3) }\right]
\label{recurrence}
\ee

\vskip 0.3 cm

 The convergence of (\ref{Konts}) was studied by Di Francesco and Itzykson 
\cite{DI}. They proved that 
$$
 A_{k}= b~ a^k ~k^{-{7\over 2}}~\left(1+O\left({1\over k}\right) \right),
~~~~~k\to \infty
$$
and numerically extimated  
$$
  a=0.138, ~~~~b=6.1~~~.
$$
We remark that the problem of the exact computation of $a$ and $b$ is open. 
 The result implies that $\varphi(\tau)$ converges in a neighborhood of 
$\tau=0$ with radius of convergence ${1\over a}$.

 We remark that as far as the Gromov-Witten invariants of genus one
  are concerned, B. Dubrovin and Y. Zhang  proved in \cite{DZ} that their 
$G$-function has the same radius of convergence of (\ref{baba}). 
Moreover, they proved the 
asymptotic formula for such invariants as conjectuder by 
Di Francesco--Itzykson. As far as I know, such a result was explained 
in lectures, but not published.

The proof of \cite{DI} is divided in two steps. The first is based on the 
  relation (\ref{recurrence}), to prove that 
$$
    A_k^{1\over k}\to a \hbox{ for } k \to \infty, ~~~~{1\over 108}
                                                              < a < {2\over 3}
$$
 $a$ is real positive because the $A_k$'s are such. It follows that  we 
can rewrite  
$$
  A_k= b a^k ~k^{\omega}~ \left(1+O\left({1\over k}\right) \right),~~~~
\omega
       \in {\bf R}
$$
The above estimate implies that $\varphi(\tau) $ has the radius of convergence 
${1\over a}$. 
The second step is the determination of $\omega$ making use of the 
differential equation (\ref{WDVVK}). Let's write 
$$
   A_k:=C_k~a^k
$$
$$
   \Phi(X)= \sum_{k=1}^{\infty} ~A_k ~e^{kX}= \sum_{k=1}^{\infty} ~C_k 
                                             ~e^{k(X-X_0)},~~~~X_0:=\ln{1\over
                                             a}
$$
The inequality ${1\over 108} < a < {2\over 3}$ implies that $X_0>0$. 
The series converges at least for  $\Re X <X_0$.  
To determine $\omega$ we divide $\Phi(X)$ into a regular part at $X_0$ and 
a singular one. Namely
$$
  \Phi(X)= \sum_{k=0}^{\infty} d_k (X-X_0)^k+ 
           (X-X_0)^{\gamma}~ \sum_{k=0}^{\infty} e_k (X-X_0)^k,~~~~
      \gamma>0 ,~~~\gamma\not\in {\bf N},
$$
$d_k$ and $e_k$ are coefficients. By substituting into (\ref{WDVVK}) we 
see that the equation is satisfied only if  $\gamma={5\over 2}$. Namely:
$$
 \Phi(X)= d_0+d_1(X-X_0)+d_2(X-X_0)^2+e_0(X-X0)^{5\over 2}+...
$$
This implies that $\Phi(X)$,  $\Phi^{\prime}(X)$ and  $\Phi^{\prime
\prime}(X)$ exist at $X_0$ but $\Phi^{\prime\prime\prime}(X)$ diverges like
\be
  \Phi^{\prime\prime\prime}(X)\asymp {1\over \sqrt{X-X_0}},~~~~X\to X_0
\label{behaviourhh}
\ee
On the other hand $
 \Phi^{\prime\prime\prime}(X)$ behaves like the series
$$ \sum_{k=1}^{\infty}~b~k^{\omega+3}~
       e^{k(X-X_0)}, ~~~~\Re (X-X_0)<0
$$
Suppose $X\in {\bf R}$, $X<X_0$. Let us put $\Delta:=X-X_0<0$. 
The above series is 
$$
 {b\over |\Delta|^{3+\omega}}~\sum_{k=1}^{\infty} (|\Delta|k)^{3+\omega}
 e^{-|\Delta|k}
\sim {b\over |\Delta|^{3+\omega}}~\int_{0}^{\infty} dx~ x^{3+\omega} e^{-x} 
$$
It follows from (\ref{behaviourhh}) that this
  must diverge like $\Delta^{-{1\over 2}}$, and thus $\omega=-{7\over 2}$ (the
  integral remains finite). 

\vskip 0.2 cm
As a consequence of  (\ref{WDVVK}) and of the divergence of $ \Phi^{\prime
\prime\prime}(X)$ 
$$   
             27+2 \Phi^{\prime}(X_0) -3 \Phi^{\prime \prime}(X_0)
               =0
$$


\section{ The case of $CP^d$}\label{solutionQH}
The case $d=1$ is trivial,  the deformation being: 
$$ 
  F(t)={1\over 2} t_1^2 t_2+e^{t_2}
$$
For any $d\geq 2$, the deformation is given by the following solution of the
WDVV equations \cite{KM} \cite{Manin}:
$$
  F(t)=F_0(t)+\sum_{k=1}^{\infty}\left[ \sum_{n=2}^{\infty}
          \tilde{\sum_{\alpha_1,...,\alpha_n}} 
          ~{N_k(\alpha_1,...,\alpha_n)\over n!} ~t_{\alpha_1}...t_{\alpha_n}
\right]e^{kt_2}
$$
where
$$ 
     \tilde{\sum_{\alpha_1,...,\alpha_n}}:=
     \sum_{\alpha_1+...+\alpha_n=2n+d(k+1) +k-3} 
$$ 
Here $N_k(\alpha_1,...,\alpha_n)$ 
is the number of rational curves $CP^1\to CP^d$ of degree $k$
through $n$ projective subspaces of codimensions
$\alpha_1-1,...,\alpha_n-1\geq2$ in general position. 
In particular,  there is one line through
two points, then 
$$
   N_1(d+1,d+1)=1
$$
Note that in Kontsevich solution $N_k=N_k(d+1,d+1)$. 

 In flat coordinates the 
 {\it Euler vector field} is 
$$ 
   E= \sum_{\alpha \neq 2} ~(1-q_{\alpha})t^{\alpha}{\partial \over 
\partial t^{\alpha} }
+~k {\partial \over \partial t^2}
$$
$$ q_1=0,~q_2=1,~q_3=2,~...,~q_{k}=k-1
$$
and 
$$\hat{\mu}= \hbox{diag}(\mu_1,...,\mu_k)=\hbox{diag}(-{d\over
2},-{d-2 \over 2},...,{d-2 \over 2},{d \over 2}),
~~~~~\mu_{\alpha}=q_{\alpha}-{d\over 2}$$


\section{ Nature of the singular point $X_0$}\label{Nature of the singular
point}

 We are now ready to formulate the problem of the paper. We need to investigate the nature of the singularity $X_0$, namely 
whether it 
corresponds   
to the fact that two canonical coordinates $u_1$,
 $u_2$, $u_3$ merge. 
Actually, we pointed out  that the structure of the semi-simple
manifold may become singular in such points because the solutions of the
boundary value problem are meromorphic on the universal covering of ${\bf
C}^n\backslash\hbox{diagonals}$ and are multi valued if  $u_i-u_j$ ($i\neq j$) 
goes around  a loop around zero.  We will verify that actually $u_i$, $u_j$ do not merge,  
but the change of coordinates $u\mapsto t$ is singular at $X_0$. 
  In this section we restore the upper indices for the flat
coordinates $t^{\alpha}$.

The canonical coordinates can be computed from the  
intersection form. We recall that the flat metric is 
$$\eta =(\eta^{\alpha\beta}):=\pmatrix{0 & 0 & 1 \cr
                 0 & 1 & 0 \cr
                 1 & 0 & 0 \cr}
$$
The intersection form is given by the formula (\ref{dall'MSRI}):
$$
  g^{\alpha\beta}= (d+1-q_{\alpha}-q_{\beta})~\eta^{\alpha \mu} \eta^{\beta\nu}
  \partial_{\mu}
  \partial_{\nu} F+ A^{\alpha\beta}, ~~~\alpha,\beta=1,2,3,
$$
where $d=2$ and the {\it charges} are $q_1=0$, $q_2=1$, $q_3=2$. The matrix
$A^{\alpha\beta} $ appears in the action of the  Euler vector field
$$
 E:=t^1\partial_1 +3 \partial_2-t^3\partial_3
$$
on $F(t^1,t^2,t^3)$:
$$
  E(F)(t^1,t^2,t^3)=(3-d) F(t^1,t^2,t^3)+ A_{\mu \nu} t^{\mu} t^{\nu}
                   \equiv
                      F(t^1,t^2,t^3)+ 3t^1t^2
$$
Thus 
$$ 
   (A^{\alpha\beta})=(\eta^{\alpha\mu}\eta^{\beta\nu} A_{\mu \nu})
                     =\pmatrix{0 & 0 & 0 \cr
                               0 & 0 & 3 \cr
                               0 & 3 & 0 \cr}
$$
 After the above preliminaries, we are able to compute the intersection form:
$$
(g^{\alpha\beta})= \pmatrix{
                            {3\over [t^3]^3}[2\Phi - 9 \Phi^{\prime}
                            +9\Phi^{\prime\prime}  ]
                          &
                             {2\over [t^3]^2}[3\Phi^{\prime\prime} -
                            \Phi^{\prime} ] 
 &
             t^1  \cr\cr
        {2\over [t^3]^2}[3\Phi^{\prime\prime} -
                            \Phi^{\prime} ] 
&
 t^1+{1\over t^3} \Phi^{\prime\prime} 
&
   3
\cr\cr
t^1
&
3
&
-t^3
\cr
}
$$
The canonical coordinates are roots of
$$
 \det((g^{\alpha\beta}-u \eta)=0
$$
This is the polynomial
$$
u^3- \left(3t^1+{1\over t^3} \Phi^{\prime\prime}\right)~u^2
-\left(
-3[t^1]^2-2{t^1\over t^3} \Phi^{\prime\prime}+{1\over [t^3]^2} 
(9\Phi^{\prime\prime}+15\Phi^{\prime}-6\Phi)
\right)        ~u    
+P(t,\Phi)$$
where
$$
 P(t,\Phi)={1\over [t^3]^3}\left( -9t^1t^3 \Phi^{\prime\prime}+243 \Phi^{\prime\prime}- 
243\Phi^{\prime}+6\Phi
 \Phi^{\prime} +\right.$$
$$\left. -9( \Phi^{\prime\prime})^2 +6t^1t^3\Phi+[t^1]^2[t^3]^2
 \Phi^{\prime\prime} -3  \Phi^{\prime}\Phi^{\prime\prime}+[t^1]^3[t^3]^3 -4
 ( \Phi^{\prime})^2 +54 \Phi-15 t^1t^3\Phi^{\prime}\right)
$$
It follows that 
$$
   u_i(t^1,t^3,X)=t^1+{1\over t^3} {\cal V}_i(X)
$$
$ {\cal V}_i(X)$ depends on $X$ through $\Phi(X)$ and derivatives. We also
observe that 
$$
  u_1+u_2+u_3=3t^1+{1\over t^3}\Phi^{\prime\prime}(X)
$$

\vskip 0.2 cm
As a first step, we 
 verify numerically that $u_i\neq u_j$ for $i\neq j$ at $X=X_0$. In order to
do this we need to compute $\Phi(X_0)$, $\Phi^{\prime}(X_0)$ ,
$\Phi^{\prime\prime}(X_0)$ in the following approximation
$$
  \Phi(X_0) \cong \sum_{k=1}^{N}~A_k~{1\over a^k},~~~
  \Phi^{\prime}(X_0) \cong \sum_{k=1}^{N}~k~A_k~{1\over a^k},~~~
~~~\Phi^{\prime\prime}(X_0) \cong \sum_{k=1}^{N}~k^2~A_k~{1\over a^k},
$$
We fixed $N=1000$ and we computed the $A_k$,
 $k=1,2,...,1000$  exactly using
 the relation (\ref{recurrence}). Then we computed $a$ and $b$ by
 the least squares method. For large $k$, say for $k\geq N_0$, we assumed that 
\be
 A_k\cong b a^k k^{-{7\over 2}}
\label{linehh}
\ee
which implies
$$\ln(A_k~k^{7\over 2} )
                                \cong (\ln a)~k +\ln b
$$
The corrections to this law are $O\left({1\over k}\right)$. 
This is the line to fit the data $k^{7\over 2}A_k$. Let 
$$
 \bar{y}:={1\over N-N_0+1} \sum_{k=N_0}^N~\ln(A_k~k^{7\over 2}),~~~~
  \bar{k} :={1\over N-N_0+1}\sum_{N_0}^N~k.
$$
By the least squares method 
$$
  \ln a = {\sum_{k=N_0}^N ~(k-\bar{k}) (\ln(A_k~k^{7\over 2} )-\bar{y}) 
           \over \sum_{k=N_0}^N~(k-\bar{k})^2}, ~\hbox{ with error }
           \left({1\over \bar{k}^2}\right)
$$
$$
  \ln b = \bar{y} -(\ln a)~\bar{k,}~\hbox{ with error }
           \left({1\over \bar{k}}\right)
$$
 For $N=1000$, $A_{1000}$ is of the order $10^{-840}$. In our computation we
 set the accuracy to $890$ digits. Here is the results, for three choices of
 $N_0$. The result should improve as $N_0$ increases, since the approximation 
 (\ref{linehh}) becomes better. 
$$
  N_0=500,~~~~a= 0.138009444...,~~~b=6.02651...
$$
$$
 N_0=700,~~~~ a= 0.138009418...,~~~b=6.03047...
$$
$$
 N_0=900,~~~~ a= 0.138009415...,~~~b=6.03062...
$$
 It follows that (for $N_0=900$) 
$$
  \Phi(X_0)=4.268908...~,~~~~\Phi^{\prime}(X_0)=5.408...~,~~~~
\Phi^{\prime\prime}(X_0)=12.25... 
$$ 
With these values we find
$$   
             27+2 \Phi^{\prime}(X_0) -3 \Phi^{\prime \prime}(X_0)=1.07...,
$$
 But the above should  vanish! The reason why this does not happen 
is that $ \Phi^{\prime \prime}(X_0)= 
\sum_{k=1}^{N}~k^2~A_k~{1\over a^k}$ converges slowly. To obtain a better
approximation we compute it numerically as 
$$
    \Phi^{\prime \prime}(X_0)={1\over 3}(27+2 \Phi^{\prime }(X_0))
                             ={1\over 3}(27+2\sum_{k=1}^{N}~k~A_k~{1\over
                             a^k})=12.60...
$$
 Substituting into $g^{\alpha\beta}$ and setting $t^1=t^3=1$ we find
$$
  u_1\approx 22.25...~,~~~u_2\approx -(3.5...)-(2.29...)i~,~~~
  u_3 =\bar{u}_2,
$$
where $i=\sqrt{-1}$ and 
the bar means complex 
conjugation. Thus, with a sufficient accuracy, we have verified
 that $u_i\neq u_j$ for $i\neq j$. 

\vskip 0.3 cm
 We now 
prove that the singularity is a singularity for the change of coordinates 
$$
   (u_1,u_2,u_3)\mapsto (t^1,t^2,t^3)
$$
 We recall that 
$$
   {\partial u_1\over \partial t^{\alpha}}= 
{(\phi_0)_{i\alpha}\over (\phi_0)_{i1}} $$
This may become infinite if $(\phi_0)_{i1}=0$ for some $i$. In our case 
$$ u_1+u_2+u_3= 3 t^1+ {1\over t^3}\Phi(X)^{\prime\prime},~~~{\partial X\over
\partial 
t^1} =0,~~~ {\partial X\over \partial
t^2} =1,~~~ {\partial X\over \partial
t^3} ={3\over t^3}
$$
and 
$$
   {\partial\over \partial t^1}(u_1+u_2+u_3)= 3,$$
$$
   {\partial\over \partial t^2}(u_1+u_2+u_3)= {1\over
   t^3}\Phi(X)^{\prime\prime\prime}  ,$$
$$
   {\partial\over \partial t^3}(u_1+u_2+u_3)= -{1\over
   [t^3]^2}\Phi(X)^{\prime\prime} +{3\over [t^3]^2}\Phi(X)^{\prime\prime\prime}
    .
$$
The above proves  that the change of coordinates is singular because 
 both ${\partial\over \partial t^2}(u_1+u_2+u_3)$ and 
$ {\partial\over \partial t^3}(u_1+u_2+u_3)$ behave like 
$\Phi(X)^{\prime\prime\prime}\asymp{1\over \sqrt{X-X_0}}$ for $X\to X_0$.

\baselineskip 12pt
\vskip 1 cm
\noindent
{\it Acknowledgments.}

I thank  A. Its and P. Bleher for suggesting me to try 
 the computations of this paper and for discussions. 
I thank B. Dubrovin for introducing me to the theory of Frobenius manifolds and for discussing together
 the problem of this paper.  
The author is supported by a fellowship of  the Japan Society for the 
Promotion of Science (JSPS).



\end{document}